\documentstyle[preprint,revtex,eqsecnum]{aps}
\def\hctwo{{$H_{c2}$}}
\def\hcone{{$H_{c1}$}}
\def\r{{\bf r}}
\def\s{{\bf s}}
\def\k{{\bf k}}
\def\R{{\bf R}}
\def\v{{\bf v}}
\def\E{{\bf E}}
\def\H{{\bf H}}
\def\B{{\bf B}}
\def\A{{\bf A}}
\def\Q{{\bf Q}}
\def\J{{\bf J}}
\def\M{{\bf M}}
\def\h{{\bf h}}

\def\j{{\bf j}}
\def\sigxx{{\sigma_{xx}^{(n)}}}
\def\sigxy{{\sigma_{xy}^{(n)}}}

\def\sig{{\sigma^{(n)}}}
\def\gr{{\gamma_{1}}}
\def\gi{{\gamma_{2}}}
\def\z{{\hat{z}}}
\begin{document}
\draft
\begin{title}
Transport Properties and Fluctuations \\
 in Type II Superconductors Near $H_{c2}$
\end{title}
\author{Robert J. Troy and Alan T. Dorsey}
\begin{instit}
 Department of Physics,
 University of Virginia,\\
 McCormick Road,
 Charlottesville, Virginia 22901
\end{instit}
\begin{abstract}

We study the flux-flow Hall effect and thermomagnetic transport near the
upper critical  field \hctwo\ in extreme type-II superconductors starting
from a suitable generalization of the time dependent Ginzburg-Landau
equations.   We explicitly incorporate the effects of backflow into the
calculations of the local electric field and current, leading to a
current which is properly divergenceless.  The Hall conductivity calculated
from this current agrees with other mean-field calculations which assume
a uniform applied electric field (the Schmid-Caroli-Maki solution),
thereby vindicating these simplified treatments.
We then use these results to calculate the transverse thermomagnetic
effects (the Ettingshausen and Nernst effects).
The effects of thermal fluctuations and nonlocal elasticity of the flux lattice
are incorporated using a method recently developed
by Vecris and Pelcovits [G. Vecris and R. A. Pelcovits,
Phys. Rev. B {\bf 44}, 2767 (1991)].  Our results, taken together with those
of Vecris and Pelcovits, provide a rather complete description of the
transport properties of the flux lattice state near \hctwo, at least
within the framework of time dependent Ginzburg-Landau theory.

\end{abstract}

\narrowtext
\section{INTRODUCTION}
\label{sec:intro}

The thermodynamic and transport properties of the mixed state of
type-II superconductors
continue to attract the interest of theorists and experimentalists alike,
due in large measure to the unusual transport properties of the
high temperature superconductors.  High transition temperatures, short
coherence lengths, and large anisotropies conspire to produce enhanced
thermal fluctuations in these materials, which can significantly modify the
mean-field phase diagram;  we refer the reader to Ref. \cite{fisher91}
for a detailed discussion of these effects.  These fluctuations
are also apparent in the transport properties, as they lead to
a broadened resistive transition in the flux flow regime near
\hctwo\ (when pinning is unimportant), and to thermally assisted flux
flow at lower temperatures (but away from the putative vortex-glass
transition \cite{fisher91}).  Indeed, if we had a detailed theory
of the transport properties in the presence of fluctuations we could
in principle use this to infer properties of the equilibrium phases.
So far, most of the theoretical work on transport
properties has focused on understanding the behavior of the
longitudinal conductivity of the flux lattice.  However, it is really
the Hall effect which represents the greatest challenge to our
understanding of the dynamics of the vortex lattice in superconductors,
as evidenced by the experimental observation that the Hall conductivity
changes sign upon entering the mixed state in the high $T_{c}$
superconductors \cite{iye89,artemenko89,hagen90,hagen91,chien91,luo92},
a feature which is at odds with the classic theories of vortex motion
in superconductors \cite{bardeen65,nozieres66,vinen67}. Motivated
by these observations, in this paper we re-examine the theory
of the Hall effect in the mixed state near \hctwo\ using a variant
of the standard time dependent Ginzburg-Landau (TDGL) theory.
By incorporating the effects of thermal fluctuations, the nonlocal
elasticity of the flux lattice, and backflow, we have consolidated
the results of several previous authors into a rather complete
theory of the Hall effect near \hctwo\ (at least within the TDGL
framework).  As a byproduct, we also study
transverse thermomagnetic effects such as the Ettingshausen effect
and the Nernst effect.

As this paper is in a sense a consolidation of the results of
several different authors, it is appropriate to first briefly
review the history of the subject. Schmid \cite{schmid66} first
derived a set of TDGL equations from the microscopic Gorkov equations.
{}From these equations he was able to calculate the flux flow conductivity
both near \hctwo\ (by solving the linearized equations) and near \hcone\ (for
a single vortex). The behavior near \hctwo\ was obtained by assuming that
the applied electric field $\E$ was constant in space;  the flux lattice is
then effectively ``boosted'' by a velocity $\v= \E\times\B/B^{2}$,
with $\B$ the induction field.  Similar methods were also used by Caroli
and Maki \cite{caroli67} to study both the dirty and clean limits.
We will henceforth refer to this solution as the Schmid-Caroli-Maki solution.
Unfortunately, the local current (which includes the normal current
plus the supercurrent) obtained  using this method is not divergence free,
as was first pointed out by Thompson and Hu \cite{thompson71}.
To obtain a current with zero divergence, it is necessary to incorporate
backflow currents; however, the backflow has zero spatial average, so that
the spatially averaged conductivity calculated using this method
agrees with the Schmid-Caroli-Maki result.  These calculations
have recently been taken one step further by Vecris and Pelcovits
\cite{vecris91}, who studied the effects of the elastic fluctuations of
the flux lattice on the conductivity.  Starting from the TDGL equations,
these authors calculated the local current with backflow, and incorporated
the elastic fluctuations by using a dynamic generalization of of the
formalism developed by Brandt \cite{brandt77} for the static flux lattice.

In all of these cases the TDGL equations which were employed
had a purely real relaxation time, and therefore exhibited a type of
``particle-hole'' symmetry which leads to a Hall conductivity which
is identically zero \cite{ullah91,dorsey92a,dorsey92b}.  To obtain a
nonzero Hall conductivity one needs to generalize the TDGL equations
by allowing the relaxation time to be complex. The imaginary part of
the relaxation time might result from either considerations of Galilean
invariance \cite{dorsey92b,maki69a}, or from microscopic considerations,
such as fermi surface curvature \cite{ebisawa71}.   Maki \cite{maki69a}
and Ebisawa \cite{ebisawa72} have used  TDGL equations with a complex
relaxation time to calculate the Hall conductivity using the
Schmid-Caroli-Maki method (i.e., without backflow).  These
equations have also been used to study the fluctuation Hall effect
for temperatures $T>T_{c2}$\ \cite{ullah91,ebisawa71}. More recently,
one of us (A.T.D.) has used the generalized TDGL equations to study the
dynamics of a single vortex (i.e., for fields close to \hcone)
\cite{dorsey92b}.

In this paper we calculate the transport properties
for the mixed state of type-II superconductors starting from a set
of TDGL equations which have a complex relaxation time. The results of
this paper therefore complement the single vortex results obtained in
Ref. \cite{dorsey92b}. The paper is organized as follows.  In Sec. II
we calculate the longitudinal and Hall conductivities in mean-field theory
by explicitly including
backflow, thereby extending the work of Thompson and Hu and Vecris
and Pelcovits. One of the important results of this
section is that the backflow current, while important in ensuring that the
total current has zero divergence, does not contribute to the spatially
averaged Hall conductivity (which is of experimental relevance).
Hence, the Hall conductivity which we obtain agrees with the result
which would be obtained using the Schmid-Caroli-Maki method.
We also briefly discuss the relevance of our results to the issue
of the sign change of the Hall conductivity in the mixed state of the
high $T_{c}$ superconductors.
In Sec. III we calculate the Ettingshausen and
Nernst effects in the presence of backflow. Our derivation, which utilizes
a recently discovered ``virial theorem'' for the equilibrium Ginzburg-Landau
equations, is exact within mean-field theory. The effects of elastic
fluctuations of the flux lattice are considered in Sec. IV, which
follows the work of Vecris and Pelcovits.
We find that the amplitude fluctuations in the
flux lattice phase suppress the flux flow conductivities below their
mean-field values.  Nonlocal effects are extremely important in
setting the scale for these fluctuations.

\section{Mean Field Theory}
\label{sec:mft}

The TDGL equations consist of an equation of motion for the order
parameter $\psi$,
\begin{equation}
(\gamma_{1}+i\gamma_{2})(\partial_{t} +i\Phi)\psi
    =  ({\nabla\over\kappa}-i\A)^{2}\psi +\psi - |\psi|^{2}\psi,
\label{tdgl1}
\end{equation}
along with Amp\`{e}re's Law,
\begin{equation}
\nabla\times\h =\J,
\label{tdgl2}
\end{equation}
where $\h=\nabla\times\A$ is the local magnetic induction field.
For the current we adopt a two fluid model, so that $\J=\J_{n}+\J_{s}$.
The normal current is
\begin{equation}
\J_{n} = \sig \cdot \E,
\label{tdgl3}
\end{equation}
where the electric field is expressed in terms of the potentials as
\begin{equation}
\E = - {1\over \kappa}\nabla\Phi - \partial_{t} \A.
\label{Efield}
\end{equation}
In the normal current we include both the longitudinal and the
transverse response of the normal carriers; the ``$\cdot$'' which appears
in Eq.~(\ref{tdgl3})  indicates a tensor product, with $\sig$ the
normal state conductivity tensor,
\begin{equation}
\sig = \pmatrix{\sigxx&\sigxy\cr
               -\sigxy&\sigxx\cr}.
\end{equation}
The signs used in $\sigxy$ are appropriate for positive carriers.
The supercurrent is given by
\begin{equation}
\J_{s} = {1\over 2\kappa i} ( \psi^{*}\nabla\psi - \psi\nabla\psi^{*})
         - |\psi|^{2}\A.
\label{tdgl4}
\end{equation}
These equations are written in dimensionless variables such that lengths are
scaled by the magnetic penetration depth $\lambda$, time is scaled by
$\hbar/2m\xi^{2}$ with $\xi$ the coherence length, magnetic fields are
scaled by $\sqrt{2} H_{c}$ with $H_{c}$ the thermodynamic critical field;
$\kappa=\lambda/\xi$ is the usual Ginzburg-Landau parameter.
As an aid to the reader, important results will be explicitly expressed in
both dimensionless and conventional units.
The quantities $\gamma_{1}$ and $\gamma_{2}$ are the the real and
imaginary parts of the dimensionless order parameter relaxation time.
The scalar potential is denoted by $\Phi$; the difference between the
scalar potential and the electrochemical potential will be ignored here
(see Refs. \cite{schmid66} and \cite{vecris91} for a more extended discussion).
Since in equilibrium we will assume that we have local charge neutrality,
out of equilibrium any charge density must be $O(v)$, with $\v$ the
velocity of the vortex lattice; therefore the time variation of the charge
density is $O(v^{2})$, and will be neglected in the spirit of the linear
response calculation of this paper.  As a result, the
total current must be divergenceless; i.e., $\nabla\cdot (\J_{n} + \J_{s})=0$.

Before attempting to solve the TDGL equations, it is useful to first
simplify them somewhat.  To do this, we write the order parameter
in terms of an amplitude and a phase, $\psi(\r,t)=f(\r,t)\exp
[i\varphi(\r,t)]$.
In terms of the gauge invariant quantities
$\Q\equiv \A - \nabla\varphi/\kappa$ and $P \equiv\Phi + \partial_{t} \varphi$,
the magnetic and electric fields are
\begin{equation}
\h=\nabla\times\Q,
\label{magnetic}
\end{equation}
\begin{equation}
\E=-\frac{1}{\kappa}\nabla P - \partial_{t} \Q,
\label{electric}
\end{equation}
and the supercurrent is
\begin{equation}
\J_{s} = - f^{2}\Q.
\label{super}
\end{equation}
The real part of Eq.~(\ref{tdgl1}) is
\begin{equation}
\gr\partial_{t}f  -\gi P f
= \frac{1}{\kappa^{2}} \nabla^{2} f - Q^{2} f + f - f^{3},
\label{real}
\end{equation}
while the imaginary part is
\begin{equation}
\gi \partial_t f + \gr P f + \frac{1}{\kappa} f \nabla\cdot \Q
        + \frac{2}{\kappa} \Q \cdot \nabla f = 0,
\label{imag1}
\end{equation}
and Eqs.~(\ref{tdgl2})-(\ref{tdgl4}) become
\begin{equation}
\nabla\times\nabla\times \Q = \sig\cdot(-\frac{1}{\kappa}\nabla P
 - \partial_{t} \Q) - f^{2}\Q .
\label{vectpot}
\end{equation}
An explicit equation for $P$ may be obtained as follows. First, multiply
Eq.~(\ref{imag1}) by $f$; the gradient terms can be combined as
$\nabla\cdot\J_{s}$; then use the fact that
$\nabla\cdot\J_{s}= -\nabla\cdot\J_{n}$.  We finally obtain
\begin{equation}
\frac{1}{\kappa}\nabla\cdot[\sig\cdot(-\frac{1}{\kappa}\nabla  P
 - \partial_{t}\Q)] + \gr f^{2} P + \gi f\partial_{t}f = 0.
\label{chempot1}
\end{equation}

We begin by calculating the local electric field for the moving flux
lattice.  First, we  assume that the lattice translates uniformly,
so that $f$, $\Q$, and $P$ are only functions of ${\bf r} - \v t$.
Therefore we replace all time
derivatives in Eqs.~(\ref{real}), (\ref{vectpot}), and (\ref{chempot1})
by $-\v\cdot\nabla$. Second, as we are concerned with linear
response in this paper, we keep only terms of order the flux lattice
velocity $\v$.  In this spirit, we expand all quantities in powers
of the velocity, with the order of expansion denoted by a superscript:
$f= f^{(0)} + f^{(1)}$, $\Q = \Q^{(0)} + \Q^{(1)}$,
where $f^{(1)}$ and $\Q^{(1)}$ are $O(v)$.
Note that $P$ is $O(v)$, since the electric
field vanishes in equilibrium.  The $O(1)$ equations are simply the
equilibrium Ginzburg-Landau equations.
The electric field can therefore be written as
\begin{equation}
\E = -\v\times\h^{(0)}- \nabla\left({1\over\kappa} P - \v\cdot\Q^{(0)}\right).
\label{josephson1}
\end{equation}
Upon averaging over the volume $V$ of the sample, we find for the spatially
averaged electric field
\begin{eqnarray}
\langle \E\rangle \equiv {1\over V} \int d^{3}r\, \E (\r)
          = -\v\times \B,
\label{josephson}
\end{eqnarray}
since the average of the gradient term in Eq.~(\ref{josephson1}) can be
converted to
a surface term which vanishes at the boundaries;
$\B = \langle \h^{(0)} \rangle$ is the (equilibrium)  macroscopic
induction field.
Although  the gradient term in Eq.~(\ref{josephson1}) does not  contribute to
the spatially averaged electric field, it does contribute to the  local
electric field.  We therefore need to calculate $P/\kappa -\v\cdot\Q^{(0)}$;
an equation for this quantity follows from Eq.~(\ref{chempot1}):
\begin{eqnarray}
&&\nabla\cdot\left[\sig\cdot\nabla\left({1\over\kappa} P
       -\v\cdot\Q^{(0)}\right)\right]  - \kappa^{2}\gamma_{1}\omega^{(0)}
      \left({1\over\kappa} P -\v\cdot\Q^{(0)}\right) \nonumber \\
&&\qquad     = -\nabla\cdot\left[ \sig\cdot(\v\times\h^{(0)})
       + {\gamma_{2}\over 2}\omega^{(0)}\v\right]
      + \kappa^{2}\gamma_{1}\omega^{(0)}\v\cdot\Q^{(0)},
\label{E11}
\end{eqnarray}
where for simplicity we have introduced $\omega^{(0)}=(f^{(0)})^{2}$.
The last term on the right hand side of Eq.~(\ref{E11}) can be further
simplified by noting that from the equilibrium equations we have
$\v\cdot(\omega^{(0)}\Q^{(0)})= \nabla\cdot(\v\times\h^{(0)})$.
Simplifying the derivatives on the left hand side, we finally arrive at
\begin{equation}
\sigxx\nabla^{2}\left({1\over\kappa} P-\v\cdot\Q^{(0)}\right) -
 \kappa^{2}\gamma_{1}\omega^{(0)} \left({1\over\kappa}P -\v\cdot\Q^{(0)}\right)
       =-\nabla\cdot\j ,
\label{E2}
\end{equation}
where we have defined
\begin{equation}
   \j(\r)  \equiv \sig\cdot [\v\times\h^{(0)}(\r)]
  - \kappa^{2}\gamma_{1} \v\times\h^{(0)}(\r)
+ {\gamma_{2}\kappa\over 2}\omega^{(0)}(\r)\v.
\label{E3}
\end{equation}
Next, define the local deviations from the average equilibrium values of
the magnetic induction field and the square of the order parameter as
\begin{equation}
 \delta\h (\r) \equiv \h^{(0)}(\r) - \B ,
\label{h}
\end{equation}
\begin{equation}
\delta\omega \equiv \omega^{(0)} - \langle \omega^{(0)} \rangle,
\label{omega}
\end{equation}
so that $\langle \delta \h (\r)\rangle = \langle\delta\omega \rangle = 0$.
When these expressions are substituted into Eq.~(\ref{E3}), there will
be a constant piece which can be discarded as it will not contribute
to Eq.~(\ref{E2}).  Noting that for the equilibrium state
$\delta\h = -(\delta\omega/2\kappa)\z$, which is correct to
$O(\delta\omega^{2})$ \cite{fetter69},
we then see that it is possible to write Eq.~(\ref{E3}) in the following
form:
\begin{equation}
\j(\r) = \sig\cdot [\v\times \delta\h(\r)]
      - {\gamma_{1}\kappa\over 2}\delta\omega(\r)( \z\times\v)
      + {\gamma_{2}\kappa\over 2}\delta\omega(\r)\v.
\label{j}
\end{equation}
Eqs.~(\ref{josephson1}), (\ref{E2}) and (\ref{j}) will together determine the
local electric field, and therefore the local normal current.

The solution to Eq.~(\ref{E2}) is
\begin{eqnarray}
{1\over\kappa} P(\r)-\v\cdot\Q^{(0)}(\r) &&= -\int d^{3}r' G(\r,\r')
      \nabla'\cdot\j(\r') \nonumber \\
       && = \int d^{3}r'\,\j(\r')\cdot \nabla' G(\r,\r'),
\label{E4}
\end{eqnarray}
where $G(\r,\r')$ is the Green's function which satisfies
\begin{equation}
\left[\sigxx\nabla^{2} - \kappa^{2}\gamma_{1}\omega^{(0)}(\r)\right]
      G(\r,\r') = \delta^{(3)}(\r-\r'),
\label{green}
\end{equation}
and where in the second line of Eq.~(\ref{E4}) we have integrated by parts
and neglected a surface contribution.  With this solution it is possible to
calculate the normal current. First, take a gradient of Eq.~(\ref{green})
and multiply by the normal state conductivity tensor; after using several
vector identities, we find
\begin{eqnarray}
\sig\cdot\nabla\left({1\over\kappa}P -\v\cdot\Q_{0}\right)
      &&= \int d^{3}r'\,\j(\r')
         \sigxx\nabla\cdot\nabla'G(\r,\r')\nonumber \\
       &&\qquad  -\nabla\times \int d^{3}r'\,\j(\r')\times \left[
          \sig\cdot \nabla' G(\r,\r')\right].
\label{E5}
\end{eqnarray}
The flux lattice state is not translationally invariant, so that $G(\r,\r')$
is not a function of the coordinate difference alone.  However, sufficiently
close to \hctwo\ the order parameter amplitude is small, and we can replace
$\omega^{(0)}(\r)$ in Eq.~(\ref{green}) by its spatial average
$\langle\omega^{(0)}\rangle$, which is correct to
$O(\delta\omega)$.  Within this approximation, the
Green's function $G(\r,\r')=G(\r-\r')$;  then
$\nabla\cdot\nabla'G(\r,\r')= -\nabla^{2}G(\r-\r')$.  Combining this result
with Eq.~(\ref{green}) for the Green's function, we find that
Eq.~(\ref{E5})  becomes
\begin{eqnarray}
\sig \cdot\nabla\left({1\over\kappa}P -\v\cdot\Q^{(0)}\right) =&& - \j(\r)
 - \kappa^{2}\gamma_{1} \langle\omega^{(0)}\rangle
                     \int d^{3}r'\,G(\r-\r') \j(\r') \nonumber \\
 &&\ \  -\nabla\times \int d^{3}r'\,\j(\r')\times
              \left[ \sig\cdot \nabla' G(\r-\r')\right].
\label{E6}
\end{eqnarray}
The second term on the right hand side of Eq.~(\ref{E6}) will generally
be quite small near the transition, as $\j$ itself is $O(\delta h)$;
this is then multiplied by $\langle\omega_{0}\rangle$, rendering the second
term
doubly small near the transition.  We will therefore drop this term
in what follows.  The normal current is obtained by combining Eq.~(\ref{E6})
with the definition of the normal current, Eq.~(\ref{tdgl3}), along with
the expression for the electric field, Eq.~(\ref{josephson1});
the final result is
\begin{eqnarray}
\J_{n}(\r) =&& -\sig\cdot[\v\times\B]
        - {\gamma_{1}\kappa\over 2}\delta\omega(\r)( \z\times\v)
      + {\gamma_{2}\kappa\over 2}\delta\omega(\r)\v \nonumber \\
   && \qquad + \nabla\times \int d^{3}r'\,\j(\r')\times
               \left[ \sig\cdot \nabla' G(\r-\r')\right].
\label{E7}
\end{eqnarray}
As a check on our result, we note that the spatial average of the last
three terms on the right hand side of Eq.~(\ref{E7}) is zero, so that
$\langle\J_{n}\rangle = -\sig\cdot (\v\times\B)= \sig\cdot\langle\E\rangle$,
as required.  It is also straightforward to show that
when $\sigxy=\gamma_{2}=0$, Eq.~(\ref{E7}) reduces to the analogous results
derived by Thompson and Hu \cite{thompson71} and Vecris and Pelcovits
\cite{vecris91} in two and three dimensions, respectively.

We next calculate the linearized supercurrent for the moving flux lattice.
This calculation is most conveniently carried out in the symmetric gauge,
rather than the Landau gauge used by Thompson and Hu \cite{thompson71}.
Following Vecris and Pelcovits \cite{vecris91}, we start by using
a postulated solution for the order parameter of a uniformly translating
flux lattice,
\begin{equation}
\psi_{l}(\r(t))=\exp\left[-{B \kappa\over 4}r^{2}(t)\right] g(x(t) +iy(t)),
\label{brandt}
\end{equation}
where the subscript $l$ indicates a linearized solution,
$\r (t)=\r -\v t $ is the coordinate in the moving frame, and $g(x(t) +iy(t))$
is an analytic function (to be specified later).
In the presence of an electric field there are
corrections to $\kappa$ of $O(v^{2})$, which we drop in the spirit of our
linear response calculation \cite{vecris91}.
Substituting Eq.~(\ref{brandt}) into the first TDGL equation,
Eq.~(\ref{tdgl1}), and dropping terms proportional to
$\nabla g/g$ , we find the required symmetric gauge potential near $H_{c2}$,
\begin{equation}
\A= {B\over 2}\z \times \r -{\gamma_{1}\kappa\over 2}\z \times \v
    + {\gamma_{2}\kappa\over 2}\v.
\end{equation}
We use an analytic function, $g(x(t)+iy(t))$ \cite{vecris91,brandt77},
appropriate for a translating flux-line lattice with flux lines located
at $\r_{\nu}$ and parallel to the $z$-axis,
\begin{equation}
g(x(t)+ i y(t))=\prod_{\nu =1}^{N}\{x(t)-x_{\nu}+i[y(t)-y_{\nu}]\} .
\end{equation}
This form of the order parameter has zeros at the instantaneous
vortex positions $\r_{\nu}$.  We therefore have for the square of the
amplitude of the order parameter,
\begin{equation}
\omega_{l}(\r(t)) = \exp\left[-{\kappa B\over 2} r^{2}(t)\right]
                \prod_{\nu=1}^{N} |\r(t) - \r_{\nu}|^{2},
\label{omegal}
\end{equation}
and for the phase of the order parameter
\begin{equation}
\varphi(\r(t)) = \sum_{\nu=1}^{N} \tan ^{-1} \left[ {{y(t) - y_{\nu}}
               \over {x(t)- x_{\nu}}} \right].
\label{phase}
\end{equation}
For the gauge invariant vector potential we then have
\begin{eqnarray}
\Q &&= \A - {1\over \kappa} \nabla\varphi\nonumber \\
   && = {B\over 2}\z \times \r -{\gamma_{1}\kappa\over 2}\z \times \v
 + {\gamma_{2}\kappa\over 2}\v - {1\over \kappa} \sum_{\nu=1}^{N}
  { {\z\times [\r(t) - \r_{\nu}]}\over {|\r(t) - \r_{\nu}|^{2}}}\nonumber \\
  &&=  -{\z \times \nabla\omega_{l}\over 2\kappa\omega_{l}}
   -{\gamma_{1}\kappa\over 2}\z \times \v +{\gamma_{2}\kappa\over 2}\v.
\label{Q}
\end{eqnarray}
The linearized supercurrent $\J_{s}=-\omega \Q$ is therefore
\begin{equation}
\J_{s} = \J_{s}^{(0)} + {\gamma_{1}\kappa\over 2}\omega_{l}(\z \times \v)
      - {\gamma_{2}\kappa\over 2}\omega_{l}\v.
\label{Js}
\end{equation}
where $\J_{s}^{(0)} = \z \times \nabla\omega_{l}/ 2\kappa $ is the
uniformly translating equilibrium supercurrent.

The total current $\J=\J_{n}+\J_{s}$ is obtained by adding our expression
for the normal current, Eq.~(\ref{E7}), to our expression for the
supercurrent, Eq.~(\ref{Js}),  which we obtained above.
Using  $\v= \langle\E\rangle\times\B/B^{2}$,  we find
\begin{equation}
\J(\r) = \sigma\cdot \langle\E\rangle + \J_{s}^{(0)}
         + \nabla\times \int d^{3}r'\,\j(\r')\times
               \left[ \sig\cdot \nabla' G(\r-\r')\right],
\label{jtot}
\end{equation}
where the conductivity tensor is
\begin{equation}
\sigma = \pmatrix{\sigxx +{\gamma_{1}\kappa
    \langle\omega^{(0)}\rangle\over 2B}
   &\sigxy -{\gamma_{2}\kappa\langle\omega^{(0)}\rangle\over 2B}\cr
 -\sigxy +{\gamma_{2}\kappa\langle\omega^{(0)}\rangle\over 2B}
  &\sigxx+{\gamma_{1}\kappa\langle\omega^{(0)}\rangle\over 2B}\cr}.
\label{conductivity}
\end{equation}
The spatial averages of the last two terms on the right hand side of
Eq.~(\ref{jtot}) are zero, so that
$\langle \J\rangle = \sigma\cdot\langle \E \rangle$; we also have
$\nabla\cdot\J=0$, since $\nabla\cdot\J_{s}^{(0)} = 0$.  We have
therefore found a current which is properly divergenceless, as required.
The various terms on the right hand side of Eq.~(\ref{jtot}) also
have simple interpretations---the first term is a uniform transport
current, the second is the uniformly translating equilibrium
supercurrent, and the last term is the backflow current.   This
form of the local current was first obtained by Thompson and
Hu \cite{thompson71} for the case $\sigxy=\gamma_{2} =0$; our result
is a generalization to the situation in which there is particle-hole
asymmetry.

In mean-field theory the Abrikosov value
for $\langle\omega^{(0)}\rangle$ is (see Ref. \cite{fetter69}, for instance)
\begin{equation}
\langle\omega^{(0)}\rangle = {m\over 2\pi \hbar e^{*}}
            {H_{c2}-B\over (2\kappa^{2}-1)\beta_{A} + 1},
\label{meanfield}
\end{equation}
where
$\beta_{A}= \langle(\omega^{(0)})^{2}\rangle/(\langle\omega^{(0)}\rangle)^{2}$,
which is $1.16$ in mean-field theory for a triangular flux lattice
\cite{fetter69}.
This leads to the following expressions for the conductivities in mean-field
theory, in conventional units:
\begin{equation}
\sigma_{xx} =\sigxx +{\gamma_{1}m\over 2\pi\hbar}{1
              \over (2\kappa^{2}-1)\beta_{A} + 1} {H_{c2}-B\over B},
\end{equation}
and
\begin{equation}
\sigma_{xy} =\sigxy  - {\gamma_{2}m\over 2\pi\hbar} {1\over
      (2\kappa^{2}-1)\beta_{A} + 1} {H_{c2}-B\over B},
\end{equation}
Notice that the conductivities
have contributions from both the normal carriers and from the vortex motion.
The real part of the order parameter relaxation time, $\gamma_{1}$, is
always positive, so that this  contribution is additive for the
longitudinal conductivity.  However, the sign of $\gamma_{2}$ is
most likely determined by microscopic considerations \cite{ebisawa71};
if  $\gamma_{2} >0$, then it
is possible for the Hall conductivity to change sign in the mixed state.
Further microscopic calculations are needed to determined if this is
the source of the sign change which has been observed in the high
$T_{c}$ superconductors.

It is also possible to calculate the corrections to the local magnetic
field for a moving flux lattice.  We can do this by expressing the
local current as a curl:
\begin{eqnarray}
\nabla\times \h(\r)&&= \J(\r)\nonumber\\
    && = - \nabla\times\left[\z\, (\J_{t}\times\z)\cdot\r \right]
   - \nabla\times\left[\z\, \omega^{(0)}(\r)/  2\kappa\right] \nonumber \\
    &&\quad  + \nabla\times \int d^{3}r'\,\j(\r')\times
               \left[ \sig\cdot \nabla' G(\r-\r')\right],
\label{hlocal1}
\end{eqnarray}
where $\J_{t} = \sigma\cdot\langle\E\rangle$ is the uniform transport current.
Integrating, we obtain [$\h(\r) = h(\r) \z$]:
\begin{equation}
h(\r) = h^{(0)}(\r) -  (\J_{t}\times\z)\cdot\r  + \int d^{3}r'\,\j(\r')\times
               \left[ \sig\cdot \nabla' G(\r-\r')\right],
\label{hlocal2}
\end{equation}
where $h^{(0)} = B - \delta\omega/2\kappa$ is the equilbrium local
magnetic field.  This is a generalization to the particle-hole
asymmetric case of the result of Vecris and Pelcovits (see Eq.~(2.23) of
Ref.~\cite{vecris91}).  As noted by these authors, the second term
in Eq.~(\ref{hlocal2}), which grows linearly with distance within the
sample, is typical of magnetostatics problems in the presence of
a uniform current density.

To summarize our results so far, we have explicitly calculated the total
current for a moving flux lattice starting from the generalized TDGL equations.
This current has zero divergence; however, the conductivities calculated
from this current are identical to those which would be obtained by
using the Schmid-Caroli-Maki solution.

\section{Thermal Transport}
\label{trans}

The moving flux lattice not only produces dissipation but also
transports energy, in a direction parallel to its velocity.  In order
to calculate the transported energy in the mixed state, we start
from the expression for the energy current which is due to Schmid
\cite{schmid66}:
\begin{equation}
\J^{h} = 2\E\times\h -2\E\times\B +2 \left[\left({\nabla\over\kappa}-i\A\right)
     \psi (\partial_{{\it t}} - i\Phi)\psi^{*} + c.c.\right],
\label{heat1}
\end{equation}
where $\J^{h}$ is in units of $(H_{c}^{2}/4\pi)(\hbar/2m)(\kappa^{2}/\lambda )$
(i.e., units of energy per unit volume times velocity).
The second term, which was not considered by Schmid, is necessary in order
to subtract out the contribution from the uniform background
field $\B = \langle\h\rangle$ \cite{huebener92}.
This expression may be greatly simplified by using a sequence of
transformations which were used in Ref.~\cite{dorsey92b}; these are
reproduced here for completeness. First, in terms of the potentials
$P$ and $\Q$, along with the order parameter amplitude $f$,
to $O(v)$ Eq.~(\ref{heat1}) becomes
\begin{eqnarray}
\J^{h} &&= 2( -\frac{1}{\kappa}\nabla P +
 \v\cdot\nabla \Q^{(0)})\times (\nabla\times \Q^{(0)})- 2\E\times\B\nonumber \\
 &&\qquad  +\frac{2}{\kappa} \left[- \frac{1}{\kappa}(\v\cdot\nabla f^{(0)})
       (\nabla f^{(0)}) + P \Q^{(0)} (f^{(0)})^{2} \right].
\label{heat2}
\end{eqnarray}
 Using $\nabla\times\nabla\times\Q^{(0)} + (f^{(0)})^{2}\Q^{(0)}=0$,
the first and last terms in Eq.~(\ref{heat2})  may be combined:
\begin{equation}
(\nabla P)\times(\nabla\times\Q^{(0)}) +
   P\nabla\times\nabla\times\Q^{(0)}= \nabla\times(P\nabla \times\Q^{(0)}),
\label{heat3}
\end{equation}
where a vector identity has been used.  The second term on the left
hand side of Eq.~(\ref{heat2}) may be written as
\begin{equation}
(\v\cdot\nabla \Q^{(0)})\times(\nabla\times\Q^{(0)})
  = \nabla\times[(\v\cdot\Q^{(0)}) \nabla\times\Q^{(0)}] +
  \v (\nabla\times\Q^{(0)})^{2} - (\v\cdot\Q^{(0)})
     \nabla\times\nabla\times\Q^{(0)},
\label{heat4}
\end{equation}
where we have again used several vector identities.  Combining
Eqs.~(\ref{heat2})-(\ref{heat4}), we have
\begin{eqnarray}
\J^{h} &&= 2\nabla\times\left[ \left(-\frac{1}{\kappa}P +
\v\cdot\Q^{(0)}\right)
       \, \h^{(0)} \right]  -2\E\times\B \nonumber \\
    &&\qquad +2\left[\frac{1}{\kappa^{2}}(\v\cdot\nabla f^{(0)})(\nabla
f^{(0)})
   + (f^{(0)})^{2} (\v\cdot\Q^{(0)}) \Q^{(0)} +  \v (h^{(0)})^{2}\right].
\label{heat5}
\end{eqnarray}
Therefore the backflow terms only appear in the first two terms on the
right hand side of Eq.~(\ref{heat5}).  However, when we calculate the
spatially averaged heat current, the first term on the right hand side
of Eq.~(\ref{heat5}) can be converted into a surface term, which
vanishes;  the second term yields $-2\langle\E\rangle\times\B=-2B^{2}\v$;
therefore the backflow corrections do not enter into the calculation
of the spatially averaged energy current.  After performing the spatial
average, we find that $\langle\J^{h}\rangle = n U_{\phi}\v$, where
$n = \kappa B/2\pi$ is the vortex density ($n= B/\phi_{0}$ in conventional
units), and where $U_{\phi}$ is the transport energy per vortex;
we have
\begin{equation}
n U_{\phi} = {1\over  V} \int\, d^{3}r  \left[ {1\over \kappa^{2}}
          (\nabla f^{(0)})^{2} + (f^{(0)})^{2} (Q^{(0)})^{2} +
          2 (h^{(0)})^{2} \right] - 2 B^{2},
\label{heat7}
\end{equation}
(a factor of $1/2$ arises from an angular average).   The first
two terms are half of the kinetic energy of the superfluid, while the
third term is the magnetic field energy.  Recently, Doria {\it et al.}
have proved a ``virial theorem'' for the equilibrium Ginzburg-Landau
equations \cite{doria89}, which shows that the integral which appears
in Eq.~(\ref{heat7}) is precisely equal to $2\H\cdot\B$.  Therefore
we find that
\begin{equation}
n U_{\phi} = -2(\B - \H)\cdot\B = -8\pi {\bf M}\cdot\B,
\label{heat8}
\end{equation}
with $\M=4\pi(\B-\H)$ the spatially averaged equilibrium magnetization
of the sample. Therefore, the transport energy in dimensionless units is
\begin{eqnarray}
 U_{\phi}&& = - {4\pi\over \kappa} 4\pi M \nonumber \\
    && = {4\pi\over \kappa} {{\kappa - H}\over {\beta_{A}(2\kappa^{2}-1)}},
\label{heat9a}
\end{eqnarray}
while in conventional units we have
\begin{eqnarray}
U_{\phi}&& = - \phi_{0} M  \nonumber \\
    && ={\phi_{0} \over 4\pi} {{H_{c2} - H}\over {\beta_{A}(2\kappa^{2}-1)}},
\label{heat9b}
\end{eqnarray}
where we have used the mean-field expressions for the magnetization
\cite{fetter69}.
Using the linearized microscopic theory near $H_{c2}$, Maki \cite{maki69b}
obtained the result $U_{\phi} = -\phi_{0} M L_{D}(t)$, where $t$ is
the reduced temperature; $L_{D}(t)\approx 1$ in the dirty limit near $H_{c2}$,
so that our results agree in this limit.
Note that our result is much more general, as the derivation did not
invoke the assumption of linearity of the order parameter,
but only the assumption of linear response in the flux lattice velocity.
Therefore our result holds for the entire mixed state,
and not just near $H_{c2}$.

The thermomagnetic transport coefficients for a superconductor in the
mixed state are discussed in Ref.~\cite{dorsey92b}, to which we refer
the reader for details.  The Nernst coefficient is defined as
$\nu=E_{y}/H (\partial T/\partial x)$, under the conditions of
$J_{x}= J_{y} = \partial T/\partial y = 0$. Introducing the transport
coefficient $\alpha_{xy}$ through
$\langle\J^{h}\rangle = \alpha_{xy} \langle\E\rangle\times\z$, then
it is possible to write the Nernst coefficient as
$\nu\approx (1/TH) (\alpha_{xy}/\sigma_{xx})$, where $\sigma_{xx}$ is the full
conductivity (including both the normal state and flux flow contributions);
but from the above discussion
we see that $\alpha_{xy}=U_{\phi}/\phi_{0}$.  We therefore find for the
Nernst coefficient,
\begin{equation}
\nu =  {1\over{\phi_{0} T H}} {U_{\phi}\over \sigma_{xx}}.
\label{heat10}
\end{equation}
The Ettingshausen coefficient is defined as
${\cal E} = (\partial T/\partial y)/HJ_{x}$ under the conditions
$J_{y}^{h} = J_{y} = \partial T/\partial x = 0$.  Using the Onsager
relations, it is possible to show that ${\cal E} = T\nu/\kappa_{xx}$,
where $\kappa_{xx}$ is the thermal conductivity.

\section{Fluctuations}
\label{fluct}

Having calculated the transport properties in the mean-field regime,
we now turn to the effect of thermal fluctuations of the flux lattice on the
transport properties.  To do this, one first assumes that the flux lines are
located at $\r_{\nu}(z)= \R_{\nu} + \s_{\nu}(z)$, where $\{\R_{\nu}\}$ are the
positions of the flux lines in mean-field theory (which form
a triangular lattice), and $\{\s_{\nu}(z)\}$ are the deviations from the
mean-field positions. Expanding about the mean-field solution, and then
taking the continuum limit by replacing $\s_{\nu}(z)$ by $\s(\r)$,
the free energy becomes $F=F_{0} + F_{\rm el}$, where $F_{0}$ is the
free energy of the mean-field Abrikosov state and $F_{\rm el}$ is the
elastic free energy given by
\begin{equation}
F_{\rm el} = {1\over 2} \int {d^{3}k\over (2\pi)^{3}} \,
      s_{i}(-\k)\{ [c_{11}(\k) - c_{66}(\k)] k_{i} k_{j}
   +\delta_{ij} [ c_{66}(\k) k_{\perp}^{2} + c_{44}(\k) k_{z}^{2}]\} s_{j}(\k),
\label{elastic}
\end{equation}
where $c_{11}$, $c_{44}$, and $c_{66}$ are the uniaxial compression modulus,
tilt modulus, and shear modulus, respectively.  The derivation of the nonlocal
elastic moduli from Ginzburg-Landau theory was first carried out by
Brandt \cite{brandt77}; his results have recently been generalized to the
case of anisotropic superconductors by Houghton, Pelcovits, and
Sudb\o\ \cite{houghton89}.  The current must now be averaged with
respect to an ensemble specified by the elastic free energy;
as shown by Vecris and Pelcovits \cite{vecris91}, this is equivalent
to replacing the spatial average of the square of the mean-field order
parameter, $\langle\omega(\r)\rangle$, which appears in the expression
for the mean-field conductivities, Eq.~(\ref{conductivity}), by the spatial
and ensemble average of the square of the order parameter,
$\langle\langle\omega(\r)\rangle_{\rm th}\rangle$,
where $\langle\cdots\rangle_{\rm th}$ is the ensemble average
(we will drop the superscript on $\omega$ for simplicity).  Therefore,
in the presence of thermal fluctuations the
longitudinal conductivity (first obtained by Vecris and Pelcovits) becomes,
in conventional units,
\begin{equation}
\sigma_{xx} = \sigxx +\gamma_{1} e^{*}
           \langle\langle\omega(\r)\rangle_{\rm th}\rangle / B,
\label{sigxx}
\end{equation}
and the Hall conductivity becomes
\begin{equation}
\sigma_{xy} = \sigxy -\gamma_{2} e^{*}
    \langle\langle\omega(\r)\rangle_{\rm th}\rangle / B.
\label{sigxy}
\end{equation}

A generalized form for the square of the order parameter in the mixed state
was first suggested by Brandt \cite{brandt77} for isotropic superconductors.
Here we will use an anisotropic generalization, appropriate for superconductors
with an anisotropic effective mass tensor; this generalization
is straightforward and is implicit in the
work of Houghton, Pelcovits, and Sudb\o\ \cite{houghton89}.  The order
parameter is
\begin{equation}
\omega(\r) = N(B) \exp \left\{ - 4\pi\sum_{\nu} \int dz'
 \int {dk_{z}\over 2\pi} \int_{\rm BZ} {d^{2}k_{\perp} \over (2\pi)^{2}}
        { e^{i\k\cdot[\r - \r_{\nu}(z')]}
         \over k_{\perp}^{2} + \gamma^{2} k_{z}^{2} + k_{\psi}^{2}}\right\},
\label{omegagen1}
\end{equation}
where $N(B)$ is a magnetic field dependent normalization constant,
$\gamma^{2} = m/m_{z}$ is the mass anisotropy, with $m$ the effective
mass in the plane and $m_{z}$ the effective mass along the $z$-axis,
$\k_{\perp} = (k_{x},k_{y})$, BZ denotes an integration over the first
Brillouin zone,  $k_{\psi}^{2}= 2 (1- b)/\xi_{ab}^{2}$
in conventional units, with $\xi_{ab}$ the in-plane coherence length,
and $b=B/H_{c2}$ is a dimensionless magnetic field.
As argued by Brandt, this form of the square of the order parameter
has the proper second-order zeros at the vortex positions $\{\r_{\nu}\}$,
and reduces to the correct forms for both large and small inductions.
Expanding the exponent to first order in $\s_{\nu}$, and taking the
continuum limit, we obtain
\begin{equation}
\omega(\r) = \omega_{0} (\r) \exp \left[ i 4\pi n
 \int {dk_{z}\over 2\pi} \int_{\rm BZ}{d^{2}k_{\perp}\over (2\pi)^{2}} \,
           {\k\cdot\s(\k)\, e^{-i\k\cdot \r}
          \over k_{\perp}^{2} + \gamma^{2}k_{z}^{2} + k_{\psi}^{2} }\right],
\label{omegagen2}
\end{equation}
where $\omega_{0}(\r)$ is the square of the mean-field order parameter with
zeros at $\{\R_{\nu}\}$, and $n$ is the vortex density; the $k_{\perp}$
integration is now over a circular Brillouin zone of radius
$k_{BZ}=(4\pi n)^{1/2}= (2 b)^{1/2}/\xi_{ab}$.
Performing the thermal average, we obtain
\begin{equation}
\langle \omega (\r) \rangle _{\rm th} = \omega_{0}(\r) e^{-W},
\label{thermalave}
\end{equation}
where $W$ is a suppression factor, given by
\begin{equation}
W = {1\over 2} (4\pi n)^{2} \int {dk_{z}\over 2\pi}
    \int_{\rm BZ}{d^{2}k_{\perp}\over (2\pi)^{2}} \,
   {1 \over (k_{\perp}^{2} + \gamma^{2} k_{z}^{2} + k_{\psi}^{2} )^{2}}
   k_{i} k_{j} \langle s_{i}(\k) s_{j}(-\k) \rangle _{\rm th}.
\label{W1}
\end{equation}
The fluctuation propagator is given by
\begin{equation}
\langle s_{i}(\k) s_{j}(-\k)\rangle_{\rm th} = k_{B}T [ P_{ij}^{T} G_{T}(\k)
                  + P_{ij}^{L} G_{L}(\k)],
\label{prop1}
\end{equation}
where $P_{ij}^{T}= \delta_{ij} - k_{i}k_{j}/k_{\perp}^{2}$ and
$P_{ij}^{L}=k_{i}k_{j}/k_{\perp}^{2}$ are the transverse and longitudinal
projection operators, respectively, and where the transverse and
longitudinal propagators are given by
\begin{equation}
G_{T}(\k) = {1\over c_{66}(\k) k_{\perp}^{2} + c_{44}(\k) k_{z}^{2} }, \qquad
G_{L}(\k) = {1\over c_{11}(\k) k_{\perp}^{2} + c_{44}(\k) k_{z}^{2} }.
\label{prop2}
\end{equation}
Carrying out the implicit summation in Eq.~(\ref{W1}), we obtain
\begin{equation}
W = {1\over 2}  k_{B} T  (4\pi n )^{2} \int {dk_{z}\over 2\pi}
    \int_{\rm BZ} {d^{2}k_{\perp} \over (2\pi)^{2}} \,
   {k_{\perp}^{2} \over (k_{\perp}^{2} +\gamma^{2} k_{z}^{2}+
k_{\psi}^{2})^{2}}
    {1 \over c_{11}(\k) k_{\perp}^{2} + c_{44}(\k) k_{z}^{2} }.
\label{W2}
\end{equation}
The spatial average is trivial; our final result is
\begin{equation}
\langle\langle \omega (\r) \rangle_{\rm th}\rangle =
    \langle \omega_{0}(\r) \rangle e^{-W},
\label{W3}
\end{equation}
with $\langle \omega_{0}(\r) \rangle$ the spatial average of the square
of the mean-field order parameter given in Eq.~(\ref{meanfield}).

In order to calculate the supression factor $W$ we first rescale the momenta
by the Brillouin zone radius by introducing a dimensionless variable
$q=k/k_{\rm BZ}$.  Then we have
\begin{equation}
W ={1\over \pi} {1\over \Lambda_{T} k_{\rm BZ}} {B^{2}\over 4\pi}
      \int_{0}^{\infty} dq_{z}\, \int_{0}^{1} dq_{\perp} \,
{q_{\perp}^{3} \over [q_{\perp}^{2} +\gamma^{2} q_{z}^{2}
                                    + m_{\xi}^{2}]^{2}}
    {1 \over c_{11}({\bf q}) q_{\perp}^{2} + c_{44}({\bf q}) q_{z}^{2} },
\label{W4}
\end{equation}
where $\Lambda_{T} = \phi_{0}^{2}/16\pi^{2}k_{B} T$ is a thermal
length \cite{fisher91}, and where
$m_{\xi}^{2}=(k_{\psi}/k_{\rm BZ})^{2}=(1-b)/b$.
The nonlocal, anisotropic elastic coefficients are given by
\cite{houghton89,ikeda92}
\begin{equation}
c_{11}({\bf q}) = {B^{2}\over 4\pi} \left({k_{h}\over k_{\rm BZ}}\right)^{2}
   \left[ {q^{2} + \gamma^{2} m_{\lambda}^{2} \over (q^{2} + m_{\lambda}^{2})
   (q_{\perp}^{2} + \gamma^{2}q_{z}^{2} + \gamma^{2} m_{\lambda}^{2} )}
  - {1 \over q_{\perp}^{2} + \gamma^{2}q_{z}^{2} +m_{\xi}^{2}}\right],
\label{c11}
\end{equation}
\begin{equation}
c_{44}({\bf q}) =  {B^{2}\over 4\pi} \left({k_{h}\over k_{\rm BZ}}\right)^{2}
   \gamma^{2} \left [ 1 + {1\over q_{\perp}^{2}
                + \gamma^{2} q_{z}^{2} + \gamma^{2} m_{\lambda}^{2}} \right],
\label{c44}
\end{equation}
with $k_{h}^{2} = (1-b)/\lambda^{2}$, and
$m_{\lambda}^{2}=(k_{h}/k_{\rm BZ})^{2}=(1-b)/2b\kappa^{2}$.
The integral is rather formidable, and we have not succeeded in evaluating it
in a general form.  However, progress is possible if we consider the
limit $m_{\lambda}^{2} \ll 1$, i.e., $1/(2\kappa^{2})\ll b/(1-b)$,
which is easily satisfied for most of the mixed state in the high $T_{c}$
superconductors.
In this limit it is possible to take to nonlocal limit of the elastic
coefficients, i.e., set $m_{\lambda}^{2}=0$ in the expressions for $c_{11}$
and $c_{44}$, Eqs.~(\ref{c11}) and (\ref{c44}).  In this limit the
anisotropy factor $\gamma$ only enters when multiplied by $q_{z}$,
so it may be scaled out of the integral.  The integral on $q_{z}$
can be performed analytically, although the result is quite complicated.
The remaining $q_{\perp}$ integral can then be performed in the limits
$m_{\xi}^{2}\gg 1$ or $m_{\xi}\ll 1$. We will consider these cases in turn.

(i)  $m_{\xi}^{2}\gg 1$; i.e., $1/(2\kappa^{2})\ll b/(1-b)\ll 1$ .
In this limit, which applies to most of the mixed state,  we obtain
\begin{eqnarray}
W &&= {\kappa^{2} \over 4}\left({m_{z}\over m}\right)^{1/2}
         {1\over \Lambda_{T} k_{\rm BZ}}
          \left( {k_{\rm BZ}\over k_{\psi}}\right)^{6}\nonumber \\
  &&\approx {\sqrt{2} \over 8}\left({2\pi\epsilon m_{z}\over m}\right)^{1/2}
        {1\over (1-t)^{1/2}} {b^{5/2} \over (1-b)^{3}},
\label{finalW1}
\end{eqnarray}
where we have used $\xi = \xi_{ab}(0) ( 1 - t)^{-1/2}$ with $t=T/T_{c}$;
$\epsilon$ is the Ginzburg criterion parameter,
given by \cite{houghton89}
\begin{equation}
\epsilon = {1\over 2\pi} \left({\kappa^{2} \xi_{ab}(0)
                         \over \Lambda_{T_{c}}}\right)^{2}
     = {16 \pi^{3} \kappa^{4} (k_{B} T_{c})^{2} \over \phi_{0}^{3} H_{c2}(0)} .
\label{ginzburg}
\end{equation}
The Ginzburg parameter measures the relative strength of thermal fluctuations;
the high $T_{c}$ superconductors typically have
$(\epsilon m_{z}/m)^{1/2}$ of $O(1)$ or larger, an enhancement of several
orders of magnitude over the low $T_{c}$ superconductors
\cite{fisher91,houghton89}.  These fluctuations lead to a reduction of the
conductivity and a significant rounding of the resistive transition in
a magnetic field.

(ii)  $m_{\xi}^{2}\ll 1$; i.e., $1/(2\kappa^{2})\ll 1 \ll b/(1-b)$ .
This is the appropriate limit for fields near the mean-field
upper critical field $H_{c2}$.  In this limit we obtain
\begin{eqnarray}
W && = (\sqrt{2} - 1) \kappa^{2}\left({m_{z}\over m}\right)^{1/2}
   {1\over \Lambda_{T} k_{\rm BZ}}
    \left( {k_{\rm BZ}\over k_{\psi}}\right)^{3}\nonumber \\
&&\approx  {2-\sqrt{2}\over 2}\left({2\pi\epsilon m_{z}\over m}\right)^{1/2}
       {1\over (1-t)^{1/2}} {b \over (1-b)^{3/2}}.
\label{finalW2}
\end{eqnarray}
Again, we note the appearance of the Ginzburg parameter $\epsilon$.
We would also like to point out that $W$ exhibits an interesting scaling
behavior in this limit; i.e.,
\begin{equation}
W(B,T)  = {\cal F} \left[ \left({\Lambda_{T} \gamma
         \over \kappa^{2}\xi_{ab} b}\right)^{2/3} (1-b) \right],
\label{scaling}
\end{equation}
where the scaling function ${\cal F}(x) = 0.29\, x^{-3/2}$. Therefore the
conductivities near $H_{c2}$ exhibit a scaling behavior
quite similiar to the scaling behavior which is inherent in transport
calculations which use the lowest Landau-level Hartree approximation
\cite{ullah91} or extensions thereof \cite{ikeda89}.  This scaling
behavior has recently been observed in measurements on the high
$T_{c}$ superconductors \cite{welp91}.

The importance of nonlocal effects in setting the scale for thermal
fluctuations which may melt the flux lattice was first pointed out by
Brandt \cite{brandt89}.  This can also be observed in the amplitude
fluctuations.  In the isotropic
limit ($\gamma =1$) the  suppression factor in the local limit is easily
calculated; we find
\begin{equation}
W_{\rm local} = {\sqrt{2}\over 12} {\xi_{ab} \over \Lambda_{T}}
              {b^{3/2} \over (1-b)^{2}},
\label{Wlocal}
\end{equation}
which lacks the important factor of $\kappa^{2}$ which appears in the nonlocal
expression.

There are several features of our result for the thermally averaged order
parameter which are noteworthy.  First, the fluctuations {\it suppress}
the conductivity below the mean-field value; this should be contrasted with the
behavior above the mean-field $H_{c2}$, where fluctuations enhance the
conductivity above the normal state value. Second, notice that the
integral in
Eq.~(\ref{W2}) is infrared convergent; there are no divergences in the
amplitude fluctuations which we consider here \cite{maki71}.
This is in contrast to the
phase fluctuations, which diverge with the system size
\cite{houghton90,moore92,ikeda92}.  This divergence is taken as an
indication of the absence of off-diagonal long-range order (ODLRO) in the
flux-lattice state \cite{moore92}.  The amplitude fluctuations,
while suppressing the conductivity below the mean-field value, do not
drive the flux-flow contribution to the conductivity to zero.  We therefore
have an enhanced conductivity even in the absence of ODLRO.
Third, the amplitude fluctuations are longitudinal,
unlike the phase fluctuations which are transverse.  As a result,
our expression for $W$ does not involve the shear modulus $c_{66}$.
It would then appear that the conductivities are relatively insensitive
to a vortex lattice melting transition \cite{fisher91,houghton89,nelson89},
at which the shear modulus
would be abruptly driven to zero in crossing the liquid-solid
phase boundary \cite{marchetti90}.
However, this observation may be significantly modified once we account for
vortex pinning \cite{fisher91,safar92,worthington92}.

\section{Conclusions}
\label{Con}

To summarize, we have calculated the transport coefficients in the
mixed state using a generalized TDGL theory.   Our calculations have
explicitly incorporated ``backflow'' effects, yielding a current which
is properly divergenceless.  However, the the results which we obtain
are wholly equivalent to the Schmid-Caroli-Maki
solution of the TDGL equations, since the backflow current has zero spatial
average.  Therefore, at least within the framework of TDGL theory, the
backflow currents associated with vortex motion have little bearing on the
question of the sign change of the Hall conductivity.   We also calculated the
thermomagnetic transport properties in mean-field theory, and found that
under quite general circumstances the transport energy is proportional to the
equilibrium magnetization.  Finally, we find that elastic fluctuations of
the vortex lattice tend to suppress the conductivities.  Hartree-type
approximations, which extrapolate from the high temperature phase,
generally predict an increase in the conductivities due to fluctuations
\cite{ullah91,ikeda89}.  How to reconcile these two quite different
approaches remains an open problem.

\acknowledgments

This work was supported by NSF Grant No. DMR 89-14051. A.T.D. also
gratefully acknowledges a fellowship from the Alfred P. Sloan Foundation.

\end{document}